\documentclass[final,3p,12pt]{elsarticle}

\usepackage{amssymb}

\usepackage{lineno}

\biboptions{sort&compress}

\usepackage{url}
\usepackage{verbatim}
\usepackage{hyperref}

\journal{Current Opinion in Systems Biology}

\makeatletter
\def\ps@pprintTitle{%
     \let\@oddhead\@empty
     \let\@evenhead\@empty
     \let\@oddfoot\@empty%
     \let\@evenfoot\@oddfoot}
\makeatother

\begin{document}

\begin{frontmatter}




\title{Practical Resources for Enhancing the Reproducibility of Mechanistic Modeling in Systems Biology}

\author[1,7]{Michael L. Blinov}
\author[2,7]{John H. Gennari}
\author[3,4,7]{Jonathan R. Karr}
\author[1,7]{Ion I. Moraru}
\author[5,7]{David P. Nickerson}
\author[6,7,8]{Herbert M. Sauro}

\address[1]{Center for Cell Analysis and Modeling, UConn Health, Farmington, CT, US}
\address[2]{Department of Biomedical Informatics and Medical Education, University of Washington, Seattle, WA, US}
\address[3]{Icahn Institute for Data Science and Genomic Technology, Icahn School of Medicine at Mount Sinai, New York, NY, US}
\address[4]{Department of Genetics and Genomic Sciences, Icahn School of Medicine at Mount Sinai, New York, NY, US} 
\address[5]{Auckland Bioengineering Institute, University of Auckland, Auckland, NZ}
\address[6]{Department of Bioengineering, University of Washington, Seattle, WA, US}

\address[7]{These authors contributed equally to this work}
\address[8]{Correspondence: \href{mailto:hsauro@uw.edu}{hsauro@uw.edu}}

\begin{abstract}
Although reproducibility is a core tenet of the scientific method, it remains challenging to reproduce many results. Surprisingly, this also holds true for computational results in domains such as systems biology where there have been extensive standardization efforts. For example, Tiwari et al. recently found that they could only repeat 50\% of published simulation results in systems biology. Toward improving the reproducibility of computational systems research, we identified several resources that investigators can leverage to make their research more accessible, executable, and comprehensible by others. In particular, we identified several domain standards and curation services, as well as powerful approaches pioneered by the software engineering industry that we believe many investigators could adopt. Together, we believe these approaches could substantially enhance the reproducibility of systems biology research. In turn, we believe enhanced reproducibility would accelerate the development of more sophisticated models that could inform precision medicine and synthetic biology.
\end{abstract}


\end{frontmatter}


\section{Introduction}
\label{S:1}

One of the central pillars of the scientific method is \emph{reproducibility}~\cite{Bacon:1267}, the ability of independent researchers to \emph{reproduce} results {\em de novo} without the aid of their original investigators or their hardware and software. We believe such reproducibility would accelerate computational science by making it easier for peer reviewers to quality control reported results and by making it easier for investigators build upon reported results. In particular, integrating models of multiple biological subsystems into more comprehensive models of entire cells and organisms will only be feasible if there are high-quality components that are accessible and reusable.

While attaining this degree of reproducibility is challenging, we believe that computational scientists have access to all of the raw ingredients needed to report \emph{repeatable} results that other researchers can recreate using the same data files, software tools, and a similar computational environment. For example, tools such as Java can execute code across multiple platforms, and repositories such as Docker Hub, GitHub, and Zenodo~\cite{peters2017zenodo} make it easy to share data, code, and computational environments. 



Despite the importance of repeatability, many computational results are not repeatable. For example, Tiwari et al. recently found that they could only repeat 50\% of simulation results reported in systems biology \citet{tiwari2021reproducibility}. Similar concerns have been reported in bioinformatics~\cite{hothorn2011case} and other computational domains~\cite{national2019reproducibility, haibe2020transparency, mckinney2020reply}. One artificial intelligence researcher was so concerned that they created a web site called `Papers Without Code' (\url{https://www.paperswithoutcode.com/}) to enable investigators to report irrepeatable papers. (The same researcher also created a complimentary site called `Papers With Code' (\url{https://paperswithcode.com/}), to highlight repeatable research.) Taken together, poor repeatability is currently an endemic problem to computational research.

\begin{figure}[t]
    \centering
    \includegraphics[scale=0.6]{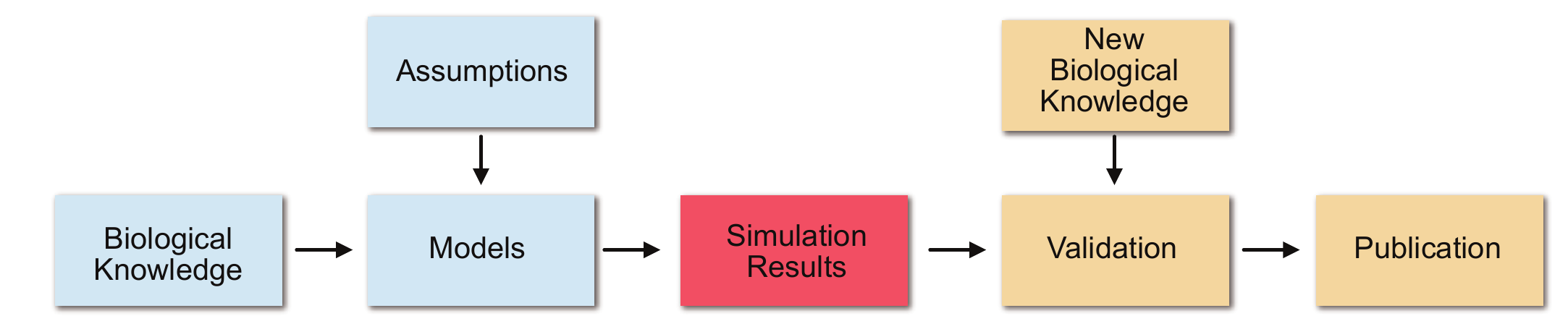}
    \caption{Typical steps for generating simulation results in systems biology. For results to be repeatable, each of these steps must be repeatable.}
    \label{fig:simpleWorkflow}
\end{figure}

In order for a computational experiment to be repeatable, each of its steps must be repeatable. Figure~\ref{fig:simpleWorkflow} illustrates the typical steps of a simulation experiment in systems biology. We and others have extensively reviewed the technical and cultural challenges to repeating each of these steps \cite{porubsky2020best, porubsky2020publishing}.

Here, we outline the key bottlenecks to the repeatability of simulations of mechanistic models in systems biology and provide practical recommendations for increasing their repeatability. In many cases, we recommend that investigators adopt practices and resources that are already widely used by the software engineering industry~\cite{hellerstein2019recent, Scholzel2020}.

\section{Toward Repeatability and Reproducibility}
To repeat a study, a researcher first needs access to the same data and software that was used in the original study, ideally in popular formats that many researchers can utilize, and with detailed metadata about how to use it, its meaning, and its provenance~\cite{wilkinson2016fair}. Access to identical hardware is typically not necessary because most researchers use similar hardware thanks to substantial standardization among computer hardware manufacturers. Access to the same operating system is also typically not needed due to extensive efforts to standardize the execution of most popular programming languages across platforms. When a specific operating system is needed, we encourage investigators to use containerization and virtualization technologies.


\subsection{Formats for Experimental Data and Knowledge}
Because computational models in systems biology are typically based on experimental data and knowledge, the ability to repeat simulation results requires this information to be published, ideally in a form that is both readily understandable by other investigators and machine-readable. In our opinion, there are many good formats for sharing different types of data and knowledge. For example, the XLSX format is often a good choice for small data sets because most researchers are familiar with the format and XLSX files are readily readable by both humans and machines. The HDF5 format~\cite{folk2011overview} can be a good choice for larger data sets because it can capture multidimensional data and metadata and is supported by most popular programming languages. Efforts to develop more expressive and higher performance formats for data include xarray \cite{hoyer2017xarray} and Zarr (\url{https://zarr.readthedocs.io/}). BioPAX is an excellent format for capturing knowledge about biochemical pathways \cite{demir2010biopax}. Whichever format investigators choose, it is critical to include metadata about the semantic meaning and provenance of this information, including information about how the data was processed and the assumptions the processing employed.

\subsection{Formats for Models}
Currently, models in systems biology are described using a broad range of formats including textual descriptions of equations in articles and supplementary materials; code for programming languages such as C/C++, Java, MATLAB, Python, and R; and domain-specific formats such as CellML~\cite{cuellar2003overview} and the Systems Biology Markup Language (SBML)~\cite{keating2020sbml}.

While pioneering studies may need to use custom formats to describe innovative experiments, this broad range of formats frustrates peer reviewers' abilities to review models and complicates investigators' abilities to reuse models. To facilitate quality control and collaboration, where possible, we recommend that investigators embrace commonly used domain-specific formats such as CellML and SBML.

The benefits of using domain-specific abstractions such as CellML and SBML are exemplified by the \TeX/\LaTeX\ document preparation system initially developed by Knuth in the late 1970's~\cite{knuth1984texbook} . Although Knuth's original implementation of \TeX\ is no longer used, \TeX\ files created over 40 years ago remain recompilable and editable today in large part because Knuth abstracted the description of \TeX\ files from the software tools used to render them into documents and because Knuth developed standards both for \TeX\ files and the documents they specify~\cite{dviStandard}. These abstractions have enabled developers to create and maintain newer \TeX\ compilers.

Some of the benefits of domain-specific formats are the abilities to develop domain-specific tools for validating models and to exchange models among software tools. For example, libSBML~\cite{bornstein2008libsbml} provides methods for validating models. Importantly, the ability to exchange models among tools ensures that modelers will be able to reuse models into the future, even if the original software tool used to develop the model is no longer available. For example, models developed in the 2000's with tools such as Gepasi~\cite{Mendes1993} and Jarnac~\cite{SauroJarnac} that were exported to SBML can still be executed with the modern tools catalogued at \url{http://sbml.org} such as COPASI~\cite{Bergmann2017}, Tellurium~\cite{Choi2018}, and Virtual Cell~\cite{loew2001virtual}.

Irrespective of which formats investigators choose to use, it is important to include metadata that enables others to evaluate and reuse models. At a minimum, models should include information about how to execute the model, the meaning of each variable and equation, and how the model was constructed and validated~\cite{le2005minimum, mulugeta2018credibility, erdemir2020credible}. Where possible, this information should be provided in a structured form that is understandable by machines. For example, the Simulation Experiment Description Markup Language (SED-ML)~\cite{Waltemath2011} and the Kinetic Simulation Algorithm Ontology (KiSAO)~\cite{courtot2011controlled} can often be used to describe simulations of models. When SED-ML cannot be used, instructions for simulating models can be provided as scripts or workflows. PEtab \cite{schmiester2021petab} is an emerging standard for capturing how models are calibrated. In many cases, modelers can use tools such as SBMLsqueezer~\cite{drager2015sbmlsqueezer} and SemGen~\cite{neal2019semgen} and ontologies such as the Systems Biology Ontology (SBO)~\cite{courtot2011controlled} and the Ontology of Physics for Biology (OPB)~\cite{cook2013ontology} to concretely describe the meaning of the biology represented by models. The Systems Biology Graphical Notation (SBGN)~\cite{Novere2009, rougny2019sbgntikz} can be used to describe diagrammatic depictions of models. New formats and ontologies must be developed to help modelers better capture the data, assumptions, and design decisions used to construct models.

\subsection{Tools for Verifying Models}
Before data and models are shared with the community, it is important for authors to verify their findings to help ensure that other investigators focus their efforts on building upon correct results. For the most part, model verification in systems biology is still {\em ad hoc} and piecemeal.

 
We believe that many investigators could enhance the quality of their data and models by adopting processes that are widely used among the software engineering industry. First, we encourage modelers to develop a list of `unit' tests that evaluate whether their simulations produce accurate results, ideally across all of the outputs of their simulations and across all of the conditions that can be captured by their simulations. Such unit tests can be organized using domain-independent frameworks such as Python's unittest module or domain-specific frameworks such as MEMOTE~\cite{lieven2020memote} and  SciUnit~\cite{omar2014collaborative}.

Second, we encourage investigators to use continuous integration (CI) systems to automate the execution of their tests each time they revise their data, models, or code~\cite{meyer2014continuous, lopez2013programming, beaulieu2017reproducibility, krafczyk2019scientific}. Used effectively, such CI systems can help investigators identify problems quickly, at a stage when they are easy to fix. Over the past five years, cloud CI systems such as CircleCI and GitHub Actions have made it easier to continuously integrate data, models, and code organized into GitHub repositories. In fact, GitHub provides Actions free for most academic projects.

\subsection{Interfaces for Repeating Simulations}
Because modern scientific data and model files are usually complex, graphical user interfaces (GUIs) are often essential to making scientific data and models reusable by investigators other than their authors. Historically, developed GUIs primarily by implementing custom desktop and web-based software tools. While this approach enables developers to create custom GUIs, the amount of work required is often impractical.

Over the past decade, Jupyter notebooks~\cite{perkel2018jupyter, Medley2018, badenhorst2019workflow, hinsen2014activepapers} have made it easier for authors to provide simple interactive, graphical windows into their data and models. These notebooks also make it easy for authors to bundle simulation experiments together with snapshots of their results and textual descriptions of their interpretation. For example, Stencila~\cite{aufreiter2018stencila} enables authors to use Jupyter-like documents to publish `live' articles~\cite{lewis2018replication} that have executable computational experiments embedded directly within them.

\subsection{Formats for Packaging Simulation Studies}
Because multiple files are often required to repeat a simulation result, it is important for these files to be distributed together in a format that preserves their links. We encourage investigators to use the COMBINE archive format~\cite{Bergmann2014} to bundle such files. COMBINE archives are zip files that include manifest files that describe their contents. The COMBINE archive format is supported by a growing number of modeling software tools and model repositories.

\subsection{Distributing Data and Models}
Over the past decade, it has become easier to publish data, models, and the software needed to reuse them. Popular avenues for sharing data and models include supplementary materials to papers; code repositories such as GitHub; domain-independent data repositories such as figshare~\cite{Singh2011}, Dryad~\cite{white2008dryad}), Harvard Dataverse (\url{https://dataverse.harvard.edu/}), and Zenodo~\cite{peters2017zenodo}; and domain-specific repositories such as BioModels~\cite{malik2020biomodels}, OpenSeek~\cite{wittig2017data, wolstencroft2015seek}, and the Physiome Model Repository (PMR)~\cite{lloyd2008cellml, yu2011physiome}. We recommend that investigators utilize archives that provide permanent storage with persistent identifiers such as BioModels and the PMR.

Popular avenues for sharing the software needed to reuse data and models include code repositories such as GitHub and software package management systems such as CRAN and PyPI. Avenues for sharing Jupyter notebooks include Binder~\cite{ragan2018binder} and CoLab~\cite{jeon2020setup}. Recently, Docker images and image repositories such as BioContainers~\cite{da2017biocontainers} and Docker Hub have become a popular way to share software~\cite{nust2020ten}. Such images make it easier for authors to share the often complex computational environments needed to reuse scientific data and models, including the software that works directly with scientific data and models, as well as all of its dependent libraries and even the operating system that it requires. 

\subsection{Curation Services}
While we believe that authors should bear most of the responsibility for reporting repeatable results, we also believe that independent curation services and journals should also play a stronger role in helping and ensuring that authors publish repeatable results. Ultimately, we aim to encourage journals to only accept articles that meet a minimum threshold, such as providing a publicly accessible version of each simulation experiment.

Due the complexity of modern computational research and the limited resources that investigators have for peer review, we believe dedicated curation services are needed both to help investigators prepare their work for dissemination and help journals rigorously evaluate both the validity and the reusability of submitted work. For example, our Center for Reproducible Biomedical Modeling (\url{https://reproduciblebiomodels.org/}) has begun to provide \textit{PLoS Computational Biology}~\cite{papin2020improving} reports of the reproducibility of computational results submitted to the journal. These reports outline whether the results reported by the authors can be recreated and whether the authors provide sufficient instructions for others to utilize their work. Others have recently launched similar efforts with the \textit{American Journal of Political Science}~\cite{APJR_2019}, \textit{Biostatistics}~\cite{peng2009reproducible}, and \textit{Physiome}~\cite{hunter2016virtual}. Over the past few years, the \textit{Journal of Open Source Software}~\cite{smith2018journal} has also begun to provide a similar service for scientific software tools.

\section{Conclusion}

Adoption of domain-independent tools such as Docker and Jupyter notebooks and domain-specific formats such as CellML, SBML, and SED-ML over the past 20 years has advanced the repeatability of computational systems biology studies. At present, many models are publicly accessible from repositories such as BioModels and PMR, the models available from these repositories often include basic metadata about model elements, and these models can often be reused with multiple simulation software tools. AS a result, by some estimates approximately 50\% of published results can be repeated with reasonable amount of effort.

Despite this progress, computational systems biology still has a long way to go to make most models reproducible and reusable. Key gaps in our ability to produce reusable models include limited tools for capturing the data and assumptions used to build models; limited adoption of newer formats such as PEtab, SED-ML, and the COMBINE archive format for describing model calibrations, specifying simulations, and bundling entire studies; and limited adoption of structured approaches to verify models such as unit testing and continuous integration. Furthermore, our scientific culture continues to under-prioritize and under-reward reproducible research~\cite{schnell2018reproducible, samota2019knowledge, hrynaszkiewicz2019publishers, heesen2018reward, quan2017publish, cowley_journalistic_2020}.  

We believe that many of these issues can be tackled by increased adoption of community standards by existing tools and the development of new tools that fill the gaps in between them. In addition, independent curation services and public model repositories could play a vital role in teaching investigators how to conduct research more reproducibly, helping researchers prepare their work for dissemination, and helping authors and journals evaluate the reproducibility of results submitted for publication.

Finally, we feel that it is critically important to shift the culture of science to more strongly value reproducibility and reusability. Because computational systems biology is not an isolated field, this issue must be addressed systemically across science and throughout the world. This will likely require reflecting these values in higher level drivers of science such as faculty hiring and promotion decisions.

Together, increased adoption of existing domain resources, targeted development of new tools, expanded curation services, and a cultural shift toward reproducibility could substantially enhance the reproducibility and reusability of computational systems biology. In turn, more reusable scientific building blocks could dramatically accelerate systems biology and the attainment of ambitious goals such as comprehensive computational models of cells and organisms that should underpin personalized medicine in the future.



\section*{Funding}

This work was supported by National Institute for Biomedical Imaging and Bioengineering award P41GM109824 and National Science Foundation award 1933453. The content expressed here is solely the responsibility of the authors and does not necessarily represent the official views of the National Institutes of Health, the National Science Foundation, the Icahn School of Medicine at Mount Sinai, the University of Auckland, the University of Connecticut, or the University of Washington. 

\section*{Acknowledgements}

We thank the COMBINE community~\cite{waltemath2020first} for many fruitful discussions that helped spur the ideas expressed here. All authors contributed equally to this work.

\section*{Declaration of competing interest}

The authors declare that they have no known competing financial interests or personal relationships that could have appeared to influence the work reported here.


\end{document}